\documentclass[prd]{revtex4}

\draft

\newcommand{\be}{\begin{equation}}
\newcommand{\ee}{\end{equation}}
\newcommand{\ba}{\begin{eqnarray}}
\newcommand{\ea}{\end{eqnarray}}

\begin{document}

\title{Dark energy  and matter from  a
five-dimensional Chern-Simons cosmology\footnote{Dedicated to Alberto Garc\'\i a on his sixthiest birthday}}
\author{ Luis F. Urrutia}
\address{
Departamento de F\'{\i}sica de Altas Energ\'{\i}as, Instituto
de Ciencias Nucleares \\Universidad Nacional Aut\'onoma de
M\'exico, A.P. 70-543, 04510 M\'exico D.F., M\'exico.}

\begin{abstract}
A Friedman-Robertson-Walker cosmology  arising from a five-dimensional Chern-Simons (CS) theory for the group $S0(1,5)$ coupled to matter is considered as an alternative model for dark energy and matter. The four-dimensional reduction describes  an accelerating  universe  having a time dependent Newton's coupling $G$ and a positive cosmological constant. Five dimensional matter gives rise to what we interprete as four dimensional ordinary plus dark matter and a dark energy  is provided by a cosmological constant term plus a fluid arising from the CS coupling. The case of five dust is studied in  detail, leading to  acceptable limits for most of the cosmological parameters considered, in the context of an open non-flat universe. Nevertheless, a value for ${\dot G}/G $
which is two orders of magnitude higher than recent bounds is predicted.
\end{abstract}

\


\maketitle

\section{Introduction}

The recent use of type I supernovae as standard candles to measure the
expansion of the universe has led to the
conclusion that the universe is accelerating \cite{perl,riess,DET}. This requires
a dark energy with significant negative pressure \cite{NEGP}. One possibility to
account for this is based on a revival  of the cosmological constant contribution. Alternative ways of
obtaining negative pressure, in this case time dependent, include a frustrated network of topological defects
(such as strings or walls)\cite{FNTD} and an evolving scalar field, named quintessence in some cases \cite{ESCF}, for example. The dark energy fractional density is estimated in the range $0.6< \Omega_{DE}<0.8$, while the corresponding ranges for matter (ordinary plus dark) fractional density are $0.15 < \Omega_m < 0.45$ (open universe) and $0.2 < \Omega_m < 0.4$ (flat universe)\cite{HAKI}.

In this work we propose an alternative way of describing an accelerated universe, which is based on  the cosmological consequences of a five-dimensional Chern-Simons (CS) theory in
our four dimensional world.

CS gauge actions, defined in odd dimensions, arise as the
boundary terms of the Chern classes $\int_{M_{2\,n}} (F)^n$, with $F=d\, A +
A\, \wedge A$ being the curvature two-form associated to the Lie group
connection one form $A$. In fact, one has that $(F)^n= d\, \Omega_{2n-1}$,
which defines the CS action to be \cite{PREP}
\begin{equation}
S_{CS}(A)=\int_{M_{2\,n-1}}\Omega_{2n-1}.
\end{equation}
In field theory the lowest dimensional case, corresponding to $n=2$,
has been profusely studied in the literature. Two of the most relevant
examples are: (i) the explicit solution of $2+1$ dimensional gravity,
rewritten as a CS theory \cite{WITT} and (ii) the realization of fractional
statistics via a three dimensional Abelian CS field \cite{LERDA}.

The next case corresponding to $n=3$ has received less attention. Following the work in Ref. \cite{CHAM} we start from a
five-dimensional Friedman-Robertson-Walker (FRW) cosmological  model  with matter. Subsequently, four dimensional quantities are introduced, thus  providing an
observable four dimensional FRW cosmology characterized by a varying Newton' s coupling. The case of vacuum solutions has been  previously discussed in Refs. \cite{BRASIL,NOE}.
Five \cite{5DIM} and higher
dimensional \cite{HDC} cosmologies arise naturally in the study of unified
models of the interactions, such as string theories for example and have
been extensively studied in the past \cite{ACF}.
Our chosen five dimensional action includes the Einstein, the Gauss-Bonet
and the cosmological constant terms with precise relative coefficients
dictated by the CS construction. We provide a matter coupling
through a five-dimensional fluid.

The paper is organized as follows: section \ref{II} summarizes the results of Ref.\cite{CHAM} which are relevant
to our purposes. The five dimensional cosmology  is introduced in section \ref{III}, where the
corresponding equations for the scale factors of the fourth ($a(t)$) and fifth dimensions ($\sigma(t)$) are summarized from the previous work in
Refs.\cite{BRASIL,NOE}. Also, the four dimensional Newton 's coupling and cosmological constant are identified.
The restriction to the case of 5-matter with zero 5-internal pressure is introduced in section \ref{IV} and the exact solutions for the scale factors are obtained.  Section \ref{V} is devoted to the identification of the corresponding four dimensional quantities which can be subsequently probed by the observational bounds. Section {\ref VI} includes a discussion of the five-dust case (corresponding to the further restriction of zero 5-external pressure) with a comparison
with the observational data. Finally, a summary of the results is given in section \ref{VII}.

\section{The model}

\label{II}

We start from the Chern-Simons action for the gauge groups $SO(1,5)$ and $%
SO(2,4)$ in five dimensions
\begin{equation}  \label{ACS}
S_{CS}(A)=k\int_{M_5}\, tr \left( A\,(d\,A)^2 +\frac{3}{2}\,A^3\,d\,A +
\frac{3}{5} A^5 \right),
\end{equation}
where the $A=A_{\bar \mu}{}^{{\bar B}{\bar C}}\, J_{{\bar B}{\bar C}}\,
d\,x^{\bar \mu}$ is the Lie-algebra valued connection one-form. The
space-time indices are ${\bar \mu}= (\mu,\,5), \, \mu=0,1,2,3$, while the
Lie-algebra indices are ${\bar B}=(B,6), \, B=(b,5),\ b=0,1,2,3.$\,$J_{{\bar
B}{\bar C}}=-J_{{\bar C}{\bar B}}$\, are the group generators normalized to $%
tr\, \left(J_{{\bar A}{\bar B}}\,J_{{\bar C}{\bar D}}\,J_{{\bar E}{\bar F}}
\right)= \epsilon_{{\bar A}{\bar B}{\bar C}{\bar D}{\bar E}{\bar F}}$. We
have omitted the explicit wedge products in (\ref{ACS}).

The equations of
motion are
\begin{equation}  \label{EQM}
\epsilon_{{\bar A}{\bar B}{\bar C}{\bar D}{\bar C}{\bar D}{\bar E}{\bar F}%
}\, F^{{\bar C}{\bar D}}\,\wedge\,F^{{\bar E}{\bar F}}=0, \qquad F^{{\bar A}{%
\bar B}}= d\, A^{{\bar A}{\bar B}} + A^{{\bar A}{\bar C}}\,\wedge\, A_{{\bar
C}}{}^{{\bar B}},
\end{equation}
where the group indices are raised or lowered by the corresponding flat
metrics $\eta_{{\bar A}{\bar B}}={\rm diag}(-,+,+,+,+,+)$ for $SO(1,5)$ and $%
\eta_{{\bar A}{\bar B}}={\rm diag}(-,+,+,+,+,-)$ for $SO(2,4)$.
Our units are such that  the form $A^{{\bar A}{\bar B}}= A_{\bar \mu}{}^{{\bar A}{\bar B}}\, d\,
x^{\bar \mu}$ is dimensionless.

In order to recover Einstein theory of gravity we follow the work of Ref. \cite
{CHAM} and introduce the following $(0,1,2,3,5)+(6)$ splitting of the
connection $A^{{\bar B}{\bar C}}= (A^{{\ B}{C}}, \ A^{B\,6})$
\begin{equation}  \label{NV}
A^{B\,C}= {\tilde \omega}{}^{B\,C}, \qquad A^{B\,6}= \eta\, {\tilde e}^B,
\end{equation}
where, as we will see from the resulting action, ${\tilde \omega}{}^{B\,C}$
are the five-dimensional Ricci rotation coefficients and ${\tilde e}^B$ are
the funfbein one-forms. Substituting in (\ref{ACS}) one obtains
\begin{eqnarray}  \label{ACC5}
S_{CS}({\tilde \omega},\, {\tilde e})= 3\,k\,\eta\int &\epsilon_{ABCDE}&
\left( {\tilde e}^A \wedge {\tilde R}^{BC}\wedge {\tilde R}^{DE} + \frac{2}{3%
} \Sigma \ \ {\tilde e}^A \wedge {\tilde e}^B \wedge {\tilde e}^C \wedge{%
\tilde R}^{DE}\right.  \nonumber \\
&&\left. + \frac{1}{5} \Sigma^2 \ \ {\tilde e}^A \wedge {\tilde e}^B \wedge {%
\tilde e}^C \wedge {\tilde e}^D \wedge {\tilde e}^E \right),
\end{eqnarray}
where
\begin{equation}{\tilde R}^{AB}= d\, {\tilde \omega}^{AB} + {\tilde \omega}%
^{AC}\,\wedge\, {\tilde \omega}_{C}{}^{B}
\end{equation}
 is the five dimensional Riemann
tensor. In the above we have defined $%
\Sigma=\lambda\, \eta^2$, with dimensions $ 1/L^2$, where $\lambda$ is related to the signature
of the fifth index group. We have $\lambda=+1$ for $%
SO(2,4)$ and $\lambda=-1$ for $SO(1,5)$. We recognize the second term in the RHS of (\ref{ACC5}) as the
five-dimensional Einstein action. The relation
\be
2k\, \eta\,\Sigma \int \epsilon _{ABCDE}e^{A}\wedge e^{B}\wedge e^{C}\wedge R^{DE}=
12\,k\,\eta\,\Sigma \int R_5 \ \sqrt{g_5}\  d^{5}x,
\ee
leads to
\be
 \frac{1}{16\,\pi\,G_5}=12\,k\,\eta\,\Sigma,
\label{G5}
\ee
thus allowing  the identification of the five-dimensional gravitational constant $G_5$.
The third term in (\ref{ACC5}) is the cosmological
constant contribution, while the first is the five-dimensional Gauss-Bonet
term.

In terms of the new variables (\ref{NV}) the equations of motion (\ref%
{EQM}) are
\begin{equation}  \label{EQM1}
\epsilon_{ABCDE} S^{AB} \wedge S^{CD}=0,\quad \epsilon_{ABCDE} S^{AB} \wedge
T^C=0,
\end{equation}
where
\begin{equation}
S^{AB}= {\tilde R}^{AB} + {\tilde e}^A \wedge {\tilde e}^B, \qquad T^A= d {%
\tilde e}^A +{\tilde \omega}^A{}_B \wedge {\tilde e}^B.
\end{equation}
Here, the torsion $T^A$ appears as the curvature ${\tilde R}^{A\,6}$. In the
following we will consider the case of zero torsion.

A full Kaluza-Klein interpretation of this model, where the five-dimensional metric is splitted in its four-dimensional metric component plus the electromagnetic and scalar fields has been developed in Refs.\cite{NOE,MACIAS}, where non-minimal couplings to gravity together with non-linear modifications to the standard Einstein-Maxwell-dilaton theory are obtained. Also, plane-fronted gravitational and electromagnetic waves solutions  in spaces endowed with a cosmological constant term are reported in the first Ref.\cite{MACIAS}.

\section{Five-dimensional cosmology.}

\label{III}

We assume that the  action (\ref{ACC5}) is coupled to general five-dimensional matter
described by the generic fluid stress tensor
\begin{equation}  \label{5DEMT}
T_{AB}=diag(\rho_{5m},\, p_{5ext\,m},\, p_{5ext\,m},\, p_{5ext\,m},\, p_{5int\,m}).
\end{equation}
Also we consider a Fermi-Robertson-Walker (FRW) line element
\begin{equation}
ds^2=-dt^2 + a^2(t) \omega^i \omega^i + \sigma^2(t) (dx^5)^2,
\end{equation}
where we use the standard parameterization  for $\omega^i,\,
i=1,2,3$ in the possible cases $k=\pm1,\, 0$. We assume that the fifth coordinate is already compactified. As we will see, this model does not accept $\sigma = cte$ as a dynamical solution, so that we will be satisfied with obtaining solutions for $\sigma(t)$ that decrease in the cosmological time.

After a standard but tedious calculation we obtain the following equations of motion for the
scale factors $a(t),\, \sigma(t)$ \cite{BRASIL,NOE}
\begin{eqnarray}
\label{E00}
&&\left[\left(\frac{\dot a}{a} \right)^2 + \frac{\dot a}{a} \frac{\dot \sigma%
}{\sigma} + \frac{k}{a^2}\right] =\frac{8\, \pi \, G_5}{3} \rho_{5m}-\frac{1}{\Sigma}\, \left( \left(\frac{%
\dot a}{a} \right)^3 \frac{\dot \sigma}{\sigma} + \frac{k}{a^2} \frac{\dot a%
}{a} \frac{\dot \sigma}{\sigma}\right)- \Sigma\,
, \\
\label{EII}
&& \left[ \frac{\ddot \sigma}{\sigma} +2 \left( \frac{\ddot a}{a} + \frac{%
\dot \sigma}{\sigma}\frac{\dot a}{a} \right) + \frac{k}{a^2} + \left(\frac{%
\dot a}{a} \right)^2 \right] = -\,8\, \pi \, G_5\, p_{5ext\,m}-
\frac{1}{\Sigma}\,\left( 2 \frac{\ddot a}{a}%
\frac{\dot a}{a}\frac{\dot \sigma}{\sigma} + \left(\frac{\dot a}{a}
\right)^2 \frac{\ddot \sigma}{\sigma } + \frac{k}{a^2} \frac{\ddot \sigma}{%
\sigma }\right)-3 \Sigma ,  \nonumber
\\
\\
&& \left[\frac{\ddot a}{a} + \left(\frac{\dot a}{a} \right)^2 + \frac{k}{a^2}
\right] =-\, \frac{8\,
\pi \, G_5}{3}\, p_{5int\,m}-\frac{1}{\Sigma}\,\left( \left(\frac{\dot a}{a} \right)^2 \frac{%
\ddot a}{a} + \frac{k}{a^2} \frac{\ddot a}{a}\right)-\Sigma.
\label{E55}
\end{eqnarray}
It will prove convenient in the sequel to introduce the following auxiliary variables \cite{BRASIL,NOE}
\begin{equation}  \label{AUXV}
v_1=\frac{{\ddot a}}{a} + \Sigma, \quad v_2= \frac{{\ddot \sigma}}{\sigma} +
\Sigma, \quad v_3=\left(\frac{{\dot a}}{a}\right)^2 +\frac{k}{a^2} + \Sigma,
\quad v_4=\frac{\dot a}{a}\frac{\dot \sigma}{\sigma} + \Sigma.
\end{equation}
Using them, Eqs.(\ref{E00},\ref{E55},\ref{EII}) are respectively written as
\begin{eqnarray}
\label{AUXV1}
v_3\, v_4= \frac{8\,\pi\, G_5\, \Sigma}{3}\,\rho_{5m},&\quad& v_1\,v_3=-\,\frac{8\,\pi\, G_5\, \Sigma}{3}\, p_{5int\, m},\nonumber \\
2\,v_1\,v_4 + v_2\,v_3&=&- \,8\,\pi\, G_5\, \Sigma\, p_{5ext\,m}.
\end{eqnarray}
The terms in square brackets in Eqs. (\ref{E00},\ref{EII},\ref{E55}) correspond to the non-zero components $G_{00},
G_{ii}$, $G_{55}$, respectively, of the five-dimensional Einstein tensor $
G_{AB}$ which satisfy well-known Bianchi identities.
These are reflected in
the conservation equation of the effective five-dimensional energy momentum
tensor, which is identified by reading off the  terms in the RHS of the above equations. In this way we can view the CS cosmology as an standard five-dimensional FRW cosmology with a modified energy momentum tensor including matter ($m$), a cosmological term ($\Lambda$)  and  a cosmic fluid ($D$) which will be interpreted as  an additional  contribution to the dark energy.
In other words we introduce the following splitting
\begin{equation}
\label{rho5T}
\rho_{5T}= \rho_{5m}+ \rho_{5D} + \rho_{5 \Lambda}, \quad p_{5A\,
T}=p_{5A\, m}+ p_{5A\,D} + p_{5A\, \Lambda},
\end{equation}
where $A$ labels the exterior and interior components of the respective pressures. The individual contributions are
\begin{eqnarray}
&&\rho_{5\,D}=-\, \frac{3}{8\, \pi\, G_5}\, \frac{1}{\Sigma}\, \left( \left(%
\frac{\dot a}{a} \right)^3 \frac{\dot \sigma}{\sigma} + \frac{k}{a^2} \frac{%
\dot a}{a} \frac{\dot \sigma}{\sigma}\right), \qquad \rho_{5\Lambda}=-\,
\frac{3}{8\, \pi\, G_5}\, \Sigma,  \nonumber \\
&&p_{5ext\, D}= \frac{1}{8\, \pi\, G_5}\, \frac{1}{\Sigma}\,\left( 2 \frac{%
\ddot a}{a}\frac{\dot a}{a}\frac{\dot \sigma}{\sigma} + \left(\frac{\dot a}{a%
} \right)^2 \frac{\ddot \sigma}{\sigma } + \frac{k}{a^2} \frac{\ddot \sigma}{%
\sigma }\right), \qquad p_{5ext\,\Lambda}= \frac{3}{8\, \pi\, G_5} \, \Sigma,
\nonumber \\
&& p_{5int\, D}=\frac{3}{8\, \pi\, G_5} \,\frac{1}{\Sigma}\,\left( \left(\frac{%
\dot a}{a} \right)^2 \frac{\ddot a}{a} + \frac{k}{a^2} \frac{\ddot a}{a}%
\right), \qquad p_{5int\, \Lambda}=\frac{3}{8\, \pi\, G_5} \, \Sigma.
\end{eqnarray}
The conservation equation for the full effective energy-momentum tensor, which matches the
corresponding five-dimensional Bianchi identity, is
\begin{equation}  \label{TMNCON}
{\dot \rho}_{5T}   + 3\, \frac{\dot a}{a}
\left(\rho_{5T}+ p_{5ext\, T}\right) + \frac{\dot \sigma}{\sigma}\, \left(\rho_{5T}+p_{5int\, T} \right)=0.
\end{equation}
We have verified that each cosmic fluid component labeled by $\Lambda$
or $D$ satisfies the conservation Eq. (\ref{TMNCON}), in such a way that the matter contribution
also does it.

At this stage we compare the second and third pieces of the action (\ref{ACC5}) with the standard four-dimensional Einstein and cosmological constant terms, respectively,
\be
I=\int \frac{1}{16\,\pi\,G_4}\sqrt{g_4}\,d^4x\,\left( R_4 - 2\Lambda_4\right),
\ee
in order to identify the corresponding Newton coupling $G_4$ and  cosmological constant $\Lambda_4$.

The resulting identification produces
\be
\frac{1}{16\,\pi\,G_4}=12 k\, \eta\,\Sigma\, \sigma(t)\, r_5, \quad -\,\frac{2\Lambda_4}{16\,\pi\,G_4}= 72k\,\eta\, \Sigma^2 \,\sigma(t)\, r_5,
\ee
which leads to
\be
\label{4DID}
G_4(t)=\frac{G_5}{r_5\,\sigma(t)},\qquad  \Lambda_4=-3\,\Sigma=-3\lambda\eta^2,
\ee
where $r_5$ is the compact radius of the fifth dimension. That is to say, $S0(1,5), \, \lambda=-1$ corresponds to the de Sitter group,
while $S0(2,4), \, \lambda=+1$ corresponds to the anti-de Sitter group.
 Let us observe that the model
predicts a time dependent Newton coupling $G_4(t)$ but a cosmological constant $\Lambda_4$.  Also we obtain
\be
\label{GDOT}
\frac{{\dot G_4}}{G_4}=- \, \frac{{\dot \sigma}}{\sigma}\,.
\ee
We choose to parameterize the equation of state of the five-dimensional matter in the standard way
\begin{equation}
p_{5ext\, m}=\left(\frac{m_3}{3}-1 \right)\, \rho_{5m} , \qquad p_{5int\,
m}=(M-1)\, \rho_{5m}.
\end{equation}
The conservation equation (\ref{TMNCON}) leads to the general form
\begin{equation}
\label{RHO5M}
\rho_{5m}\sim \frac{1}{a^{m_3}\, \sigma^M}.
\end{equation}
The equation of state  for the cosmological constant contribution is
\begin{equation}
p_{5\Lambda}=-\, \rho_{5\Lambda},
\end{equation}
while the equation of state for the D-contribution is dependent on the equations
of motion.

The vacuum solution ($\rho_{5m}=0,\, p_{5ext\,m}=0,\, p_{5int\,m}=0$) of our model  was discussed in Refs. \cite{BRASIL,NOE}. In the case of a negative deceleration parameter, the unwanted increasing behavior $ \sigma(t)={\bar \sigma}\, cosh(\Lambda t)$ for the scale factor of the fifth dimension was obtained.

\section{The case of 5-matter with 5-external pressure }

\label{IV}

This corresponds to the case where $p_{5int}=0$, ($ M=1$), with $\rho_{5m} \neq0$ and $p_{5ext\,m}\neq0$. In order to have a non-zero value for $\rho_5$ we require both $v_3$ and $v_4$ to be simultaneously non-vanishing. This leads to
\be
v_1=0, \qquad \Rightarrow \qquad {\ddot a}+ \lambda \, \eta^2 \, a=0, \qquad \lambda= \pm 1
\ee
and the equation for $a(t)$ is completely decoupled.
In the sequel we only consider the case $\lambda= -1$, which produces a negative deceleration
parameter.
The solution is
\begin{equation}
a(t)= A\,{\rm sinh}(\eta\,t).
\end{equation}
We have
\begin{equation}
H(t):=\frac{\dot a}{a}=\eta \, {\rm coth}(\eta\,t), \qquad q(t):=-\,\frac{\ddot a \,a}{{\dot a}^2}=-\, {\rm tanh}%
^2(\eta\,t),  \label{REL}
\end{equation}
together with the relations
\begin{equation}
\eta=\sqrt{-\, q(t)}\,H(t) \quad \Rightarrow \quad \eta=\sqrt{-\, q_0}%
\,H_0, \quad q(t) <0,
\label{REL1}
\end{equation}
which allows to determine the cosmological constant in terms of the present values of the Hubble and deceleration parameters.

As seen in Eq. (\ref{REL}) the deceleration parameter $q$ is negative, result
that agrees with recent observations  which
yield an accelerated expanding universe, characterized by \cite{perl,riess}
\begin{equation}
q_{0}=-0.58{\,}^{+0.14}_{-0.12},  \quad q_{0}=-1.0\pm 0.4,   \label{desa}
\end{equation}
respectively.

The next step is to deal with the solution of $\sigma(t)$. The remaining equations reduce to
\begin{eqnarray}
v_3\, v_4= \frac{8\,\pi\, G_5\, \Sigma}{3}\, \rho_{5m},  \quad
v_2\,v_3=- \,8\,\pi\, G_5\, \Sigma\,\left(\frac{m_3}{3}-1\right)\rho_{5m},
\end{eqnarray}
which ratio leads to the following equation for $\sigma$
\be
\ddot\sigma-N\eta\,coth(\eta t)\dot\sigma-(1-N)\eta^2\sigma=0,\qquad N=3-m_3.
\ee
The above can be reduced to a Legendre equation. Imposing appropriate boundary conditions we finally obtain
\ba
&& \sigma =C_{1}(x^{2}-1)^{-\nu /2}\,P_{\nu }^{\nu -1}(x), \qquad
\nu=\frac{N}{2} \leq 0, \qquad
 x= coth(\eta t).
\ea
The condition $m_3-3>0$ is indeed adequate to a standard matter distribution.
The density $\rho_{5m}$ is given by
\begin{eqnarray}
\rho _{5m} &=&\frac{3}{8\pi G_{5}}\left( \eta ^{2}+\frac{k}{A^{2}}\right)
\frac{2^{\nu }}{\Gamma (-\nu +1))}\left[ \frac{1}{\left( x^{2}-1\right)
^{\left( \nu -3\right) /2}}\frac{1}{P_{\nu }^{\nu -1}(x)}\right].
\end{eqnarray}
The matter density is positive as expected and one can
verify that $\rho _{5m}\sim
1/(a^{m_{3}}\,\sigma )$, in accordance with the conservation law (\ref
{TMNCON}).

\section{Identification of four dimensional quantities}

\label{V}

The basic five dimensional equations for our cosmology Eqs.(\ref{E00},\ref{EII}), together with the definition (\ref{rho5T}),
can be presented in the form
\begin{eqnarray}
\label{FDQ}
\left( \frac{\dot{a}}{a}\right) ^{2}+\frac{k}{a^{2}} =\frac{8\,\pi \,G_4(t)}{3}%
\,\rho _{4eff}, \qquad
2\frac{\ddot{a}}{a}+\frac{k}{a^{2}}+\left( \frac{\dot{a}}{a}\right) ^{2}
=-8\,\pi G_4(t)\,p_{4eff},
\end{eqnarray}
in order to  identify the four dimensional effective density and pressure. They turn out to be
\begin{equation}
\rho _{4eff}=r_5\,\sigma \rho _{5T}-\frac{\dot{a}}{a}r_5\,\dot{\sigma}\frac{3}{8\,\pi
\,G_{5}}, \qquad
p_{4eff}= r_5\,\sigma p_{5ext\,T}+\frac{1}{8\,\pi \,G_{5}}r_5\,\ddot{\sigma}+%
\frac{2}{8\,\pi \,G_{5}}\frac{\dot{a}}{a}r_5\,\dot{\sigma}.
\label{R4EFF}
\end{equation}
From the above equations, together with  (\ref{E55}), one can express the five dimensional density and pressures in terms of the four dimensional effective quantities. Substituting this in the five dimensional conservation equation (\ref{TMNCON}), we obtain  the four dimensional conservation equation appropriate to a universe with varying gravitational coupling  $G_4(t)$
\begin{eqnarray}
\dot{\rho}_{4eff}+3\left( \frac{\dot{a}}{a}\right) \left( \rho
_{4eff}+p_{4eff}\right) +\frac{\dot{G_4}}{G_4}\rho _{4eff} &=&0.
\label{CONS4}
\end{eqnarray}
Indeed, equation (\ref{CONS4}) can also be obtained from  the standard FRW conservation
after the replacements
$\rho _{4}\rightarrow G_4(t)\,\rho _{4eff},\;\;p_{4}\rightarrow G_4(t)\,p_{4eff}\,\,
$are made.

Let us  emphasize that once the four dimensional interpretation in enforced, the degree of freedom corresponding to the five dimensional scale factor $\sigma(t)$ is translated into a varying  Newton coupling $G_4(t)$.

Next we provide a further interpretation of the individual pieces that make up the four dimensional effective density $\rho_{4eff}$ and  the four dimensional effective
pressure $p_{4eff}$.
In the cases of  matter  and  cosmological constant we make the choice
\begin{equation}
\label{4ID5}
\rho_{4}=  r_5\,\sigma\, \rho_{5}, \qquad p_{4}=  r_5\,\sigma\,
p_{5ext},
\end{equation}
which is motivated by the standard reinterpretation of the five-dimensional total matter in a given volume, in terms of four-dimensional quantities \cite{KOLB,FARINA,PAPA}. Also, Eq.(\ref{4ID5}) preserves the form of the equations of state, leading to
\begin{equation}
\label{STEQ4}
p_{4m}=\left(\frac{m_3}{3}-1 \right)\, \rho_{4m}, \qquad  p_{4\,\Lambda}=-\rho_{4\,\Lambda}.
\end{equation}
Nevertheless, a difference arises in the individual conservation equations satisfied by each
of these components. In fact, for the cosmological constant we have
\begin{equation}
\rho_{4\,\Lambda}=-r_5\,\sigma\,\frac{3}{8\,\pi\,G_5}\,\Sigma, \qquad p_{4\,\Lambda}=r_5\,\sigma\,\frac{3}{8\,\pi\,G_5}\,\Sigma,
\end{equation}
which satisfies  the full conservation equation (\ref{CONS4}).
In the case or matter we recall that we have chosen $m_3=3, M=1, (p_{5int m}=0)$, which together with Eq.(\ref{RHO5M}) leads to
$
\rho_{4\,m}\sim 1/a^{m_3}.
$
This implies the standard matter conservation equation
\begin{eqnarray}
\dot{\rho}_{4,m}+3\left( \frac{\dot{a}}{a}\right) \left( \rho
_{4,m}+p_{4,m}\right)&=&0,
\label{CONSM}
\end{eqnarray}
associated with the first equation of  state in (\ref{STEQ4}).

Next we consider the remaining contribution to the effective quantities labeled by the index $D$. Here we simply read off the corresponding identification and conservation properties that are necessary to maintain the four dimensional Bianchi identities induced by the LHS 's of Eqs.(\ref{FDQ}). To this end we recall the over all identifications
\begin{eqnarray}
\rho_{4\,eff}&=&\rho _{4\,m}+\rho _{4\,D}+\rho _{4\,\Lambda }=r_5\,\sigma \left( \rho
_{5m}+\rho _{5D}+\rho _{5\Lambda }-\frac{\dot{a}}{a}\frac{\dot{\sigma}}{%
\sigma }\frac{3}{8\,\pi \,G_{5}}\right),\\
p_{4eff} &=& p_{4\,m}+p_{4\,D}+p_{4\,\Lambda }
=r_5\,\sigma \left( p_{5ext\,m}+p_{5ext\,D}+p_{5ext\,\Lambda }+\frac{1%
}{8\,\pi \,G_{5}}\frac{\ddot{\sigma}}{\sigma }+\frac{2}{8\,\pi \,G_{5}}\frac{%
\dot{a}}{a}\frac{\dot{\sigma}}{\sigma }\right),
\end{eqnarray}
together with  the full conservation equation (\ref{CONS4}).
Substituting our previous definitions and individual conservation equations, we are left with the remaining identification
\begin{eqnarray}
\rho_{4\,D}&=&r_5\,\sigma \left( \rho_{5D}-\frac{\dot{a}}{a}\frac{\dot{\sigma}}{%
\sigma }\frac{3}{8\,\pi \,G_{5}}\right),\\
 p_{4\,D}
&=&r_5\,\sigma \left( p_{5ext\,D}+\frac{1%
}{8\,\pi \,G_{5}}\frac{\ddot{\sigma}}{\sigma }+\frac{2}{8\,\pi \,G_{5}}\frac{%
\dot{a}}{a}\frac{\dot{\sigma}}{\sigma }\right).
\end{eqnarray}%
which must satisfy the conservation equation
\begin{equation}
\dot{\rho}_{4D}+3\,\frac{\dot{a}}{a}\left( \rho _{4D}+p_{4D}\right) +\frac{%
\dot{G}}{G}\rho _{4D}=-\,\frac{\dot{G}}{G}\rho _{4m}.
\end{equation}
The above means that matter is a source of the fluid $D$. The possible consequences of this are not explored in this work.

Next we turn to the fractional densities. To this end we  introduce the ratios $\Omega_{4 n} $ among the different
densities $\rho_{4 n}$ and the instantaneous critical density $\rho_c(t)$. We
have
\begin{equation}  \label{DEFOM}
\Omega_{4n}(t)=\frac{\rho_n(t)}{\rho_c(t)}, \qquad \rho_C(t)= \frac{3}{8\,\pi\,G_4(t)%
}H(t)^2, \quad \rho_{C0}=0.947\times 10^{-29} \, {\rm g}/{\rm cm}^3.
\end{equation}
From the first Eq.(\ref{FDQ}) we obtain the usual result
\begin{equation}
1=\Omega_{4m} + \Omega_{4\Lambda} + \Omega_{4D}+\Omega_{4k}:=\Omega+ \Omega_{4k}, \qquad \Omega_{4k}=-\,\frac{k}{(aH)^2}\,.
\end{equation}
The  fractional density at the present time corresponding to the cosmological term  is
\be
\Omega_{4\Lambda0}=\frac{\eta^2}{H_0^2}=\frac{\Lambda}{3H_0^2}=-q_0,
\ee
independently of the behavior of $\sigma(t)$.

\section{The case of five-dust with $\lambda=-1$}

\label{VI}

In order to test some observational consequences of the model under consideration we examine in some detail the simple  case when $ \, p_{5ext\,m}=0, (m_3=3)$, $ \,  p_{5int\,
m}=0, (M=1)$. The solutions are
\begin{equation}
a(t)= A\,{\rm sinh}(\eta\,t), \qquad \sigma(t)= B\,{\rm exp}({- \eta\,t%
}).
\end{equation}
Let us remark that here we can implement a decreasing fifth scale parameter producing a crack of doom singularity at $t\rightarrow\infty$. Using the above solutions together with the first Eq.(\ref{AUXV1}) we obtain
\begin{eqnarray}
\rho_{5m}&=&\frac{3}{8\,\pi\, G_5}\left(\eta^2 + \frac{k}{A^2} \right)\frac{{\rm e}%
^{\eta\, t}}{{\rm sinh}^3(\eta\, t)},
\end{eqnarray}
where the constant term involving  $k$ can be rewritten as
\be
\frac{k}{A^2}=sinh^2(\eta\,t_0)H^2_0\,(\Omega_0-1),
\ee
and the subindex zero labels the present time. Using the relations (\ref{REL}) together with the four-dimensional identifications described in the previous section it is possible to write
all the required fractional densities in terms of $\Omega_0$ and $Q_0 :=-q(t_0)$ as
\ba
\label{OMEGAS}
\Omega_{4m0}&=& (\Omega_0-Q_0)\left(1+ \frac{1}{\sqrt{Q_0}} \right), \qquad \Omega_{4\Lambda 0}= Q_0, \qquad \nonumber \\
\Omega_{4D0}&=&-(\Omega_0-Q_0)\frac{1}{\sqrt{Q_0}}, \qquad \Omega_{k0}=1-\Omega_0.
\ea
Since the individual density contribution of the fluid D is negative for the cases when the matter density is positive, we choose to interprete it as part of the dark energy density $\Omega_{DE0}$ and define
\be
\Omega_{DE0}=\Omega_{4\Lambda0}+\Omega_{4D0}, \qquad \Omega_0= \Omega_{4m0}+\Omega_{DE0}.
\ee
In  order to make some numerical estimations with this model we consider the following range for the relevant parameters \cite{HAKI}: age of the universe: $12\,< \, t_0  \,<\, 18$ Gyr, Hubble parameter:
$60\,  < \,H_0 \,< \, 82 \,$ $\rm km/(seg \, M_{pc})$,  total density parameter : $0.85\,<\,\Omega_0\, <
\, 1.25$ \cite{BOOMERANG}, and mass density parameter: $ 0.2 < \Omega_{4m0} < 0.4$. In the sequel the numerical quantities will be assumed to have the corresponding units stated above.

For fixed $\Omega_{4m0}$, the  first expression in (\ref{OMEGAS}) produces a minimum $Q_0$ for the minimum value $\Omega_0=0.85$. This corresponds to a minimum value of $t_0$ associated to the maximum $H_0= 82$.     In fact, for a given $Q_0$, the model predicts increasing $t_0$'s for decreasing $H_0$'s, according to Eq. (\ref{REL}).

The model disfavors $\Omega_0 \geq 1$, which would require at least $t_0=21$ for $H_0= 82$, with increasing values for the smaller values the Hubble parameter. The maximum values:  $H_0=82$ together with $t_0=18$, determine  the maximum value  $Q_0=0.74$. For  $\Omega_{4m0}=0.3$ this produces the maximum value $\Omega_0=0.88$ for an open non-flat universe. On the other hand, the minimum value $\Omega_0=0.85$ produces $Q_0=0.71$ for $\Omega_{4m0}=0.3$, leading to $t_0=17.4$ for $H_0=82$. This same minimun value $\Omega_0=0.85$ with lower values of $H_0$ leads to increasing ages of the universe, reaching the upper limit for $H_0=79$. Summarizing, for $\Omega_{4m0}=0.3$ the model predicts an open non-flat universe with $0.85 < \Omega_0 <0.88$, with corresponding $0.55 < \Omega_{DE0} < 0.58$. Also we have $17.4 < t_0 < 18$;  $\, 79 < H_0 < 82$;  and
$ 0.71 < Q_0 < 0.74$. The lower value $\Omega_{4m0}=0.2$ is ruled out by the model requiring a minimum of $t_0=18.3$ for $\Omega_0=0.85$ with increasing ages for higher $\Omega_0$. The minimum allowed is $\Omega_{4m0}=0.24$. Higher values of $\Omega_{4m0}$ can be accomadated in the model. For example $\Omega_{4m0}=0.4$ requires a minimun age of $t_0= 16.8$ for $H_0=82$ and $\Omega_0=0.85$ with $Q_0=0.67$.  The maximun age here is obtained with $H_0=76.5$. The maximum value for $\Omega_0$ is 0.93.

Recalling that the  model requires a time dependent Newton coupling we now examine the known bounds on this quantity. In the case of five dust our prediction,  according to  Eq. (\ref{GDOT}),  is
\be
\frac{{\dot G}_4}{G_4}=+\, \eta=\sqrt{Q_0}\,H_0 \approx\,  7.0 \times 10^{-11} \, yr^{-1}.
\label{ETAEST}
\ee
for the averages  $Q_0=0.73$ and $H_0=80$ determined previously in the case $\Omega_{4m0}=0.3$. A comparison with one of the latest  bounds: $-3\times 10^{-13} yr^{-1}< \left({\dot G}_4/G_4\right)_0 <
 4 \times 10^{-13} yr^{-1}$ \cite{COPIETAL}, shows that  unfortunately the predicted value is too high by almost two orders of magnitude. The range of allowed values of  either $Q_0$  or $H_0$ will not allow any drastic modification of the estimation (\ref{ETAEST}) in the case of five-dust.

\section{Summary}

\label{VII}

A FRW cosmological model arising from a five-dimensional Chern-Simons theory for the group $S0(1,5)$ is considered. The resulting five dimensional equations for the corresponding scale parameters $a(t)$ and $\sigma(t)$ are exactly solved in the case of five-matter with 5-external pressure. The four-dimensional reduction describes  an accelerating  universe ($q_0 <0$) having a time dependent Newton's coupling $G_4(t)$ and a positive cosmological constant. The five dimensional matter gives rise to what we interprete as four dimensional ordinary plus dark matter $\Omega_{4m0}$ and a  dark energy  $\Omega_{DE0}=\Omega_0-\Omega_{4m0}$ is provided by the cosmological constant plus a fluid component arising from the CS coupling. The case of five dust (zero 5-external pressure) is studied in more detail, leading to a decreasing behavior of the fifth dimension scale parameter, reaching zero for infinite cosmic time. For the choices $\Omega_{4m0}=0.3$ and $79 <\,H_0\,<82 $ the  dust model predicts an open non flat universe with values: $0.85 <\Omega_0 < 0.88, \, \, 17.4 < t_0 <18 \, {\rm Gyr},\,$ and $0.71 < -q_0 < 0.74 $. Unfortunately, a value for ${\dot G}_4/G_4$
which is two orders of magnitude higher than  recent bounds is found.

Within  the framework  presented so far it remains to be examined whether the choice  $p_{5ext\, m} \neq 0$, $ (m_3 \neq 3)$, could alleviate some of the above problems. Finally, we have the open alternative to explore solutions with  $p_{5int\, m} \neq 0$, $ (M \neq 1)$, which is not considered at all in this paper. The improvement of the five-dust model over the vacuum situation provides some degree of optimism in this respect.

\end{document}